\documentstyle[prl,aps,twocolumn,floats,epsfig]{revtex}
\begin{document}
\draft
\title{Temporal correlation function in $3-D$ Turbulence.}

\author{ Anirban Sain }
\address{Department of Physics, Simon Fraser University,
Burnaby, British Columbia, Canada V5A 1S6}

\date{\today}
\maketitle

\begin{abstract}
We observe oscillatory decay in the two-point, non-equal time, velocity
correlation function  of homogeneous, isotropic 
turbulence. We found this through a direct numerical simulation (DNS) 
of the three dimensional Navier-Stokes ($3-D$ NS) equation. We give an approximate 
analytic theory which explains this oscillatory behaviour. The 
wave-number and frequency dependent effective viscosity turns out to
be complex; the imaginary part gives rise to the temporal 
oscillation. We find that, at least for the decay at short times, data 
collapse occur among the inertial range velocity wave-vector modes 
with the long time dynamic exponent $z=2/3$, but the time period of the 
temporal oscillation is not universal.

\end{abstract}

In homogeneous, isotropic turbulence the main interest is to understand
the long-range spatio-temporal correlations exhibited by the velocity
field. Towards this end one studies the scaling behaviour of the velocity
structure functions $S_p(l,t)\equiv\langle [v_i(0,0)-v_i(l,t)]^p\rangle$
with respect to $l$ and $t$. Here $v_i$ is the velocity field, $l$ and $t$ are
the spatial and temporal separations. Kolmogorov had predicted \cite{k41}
through
his dimensional analysis argument that the equal time structure function 
$S_p(l,0)=({\epsilon}l)^{p/3}$. He had assumed that in the inertial
range i.e., $\eta_d \ll l \ll L\/$ ($\eta_d\/$ and $L\/$ are, 
respectively, dissipation and forcing scales) $S_p(l,0)$ is a function
of $\epsilon$ (the mean energy dissipation rate) and $l$ only. 
But after anomalous scaling properties of $S_p(l)$ were discovered through 
experiments and simulations \cite{multi}, the importance of $L$ (which 
is also called the integral scale) has been recognised. 
Now it is known that $S_p(l)\sim l^{\zeta_p}$, when $\zeta_p$
is a monotonically increasing, convex, nonlinear function of $p$.
The negative correction to the exponent $\delta\zeta_p = \zeta_p - p/3$
has to appear as the exponent of a dimensionless quantity $(l/L)$
in order to keep the dimension of $S_p(l)$ unchanged. 
In all experiments and simulations of the $3-D$ NS equation 
$L$ is finite. Infact a grand challenge for analytic theories 
of turbulence is to show that finite limit for $\zeta_p$ exist
for $l\ll L\rightarrow \infty$. Such a scheme has been 
successfully carried out \cite{passvT} for the passive scalar 
field advected by a random velocity field in three dimensions 
(the Kraichnan model \cite{kraichnan}). It has been shown that the 
anomalous scaling exponents of the passive scalar structure functions
are independent of $L$ (as $L\rightarrow \infty$), but the amplitudes 
do depend on $L$. Given that so much effort have been made to 
understand $S_p(l,0)$, not much is known about $S_p(l,t)$ even for
small integer values of $p$ (of course $S_p(l,t)$ is more complicated than 
$S_p(l,0)$).  Literature on $S_p(l,t)$ in NS turbulence or related 
Burgers turbulence is rather sparse \cite{belinicher}. Dynamic 
renormalisation group (DRG) calculations \cite{YOrg}, one loop 
self-consistent calculations \cite{jkb2} with the randomly forced
$3-D$ NS equation suggest a long time dynamic exponent $z=2/3$.
In Ref\cite{jkb2} large wavenumber limit of the velocity correlation
function was also explored assuming dynamic scaling hypothesis to
be valid. In experiments with a mean flow, because of large scale 
background velocity, one expects to measure $z=1$. But even with zero mean 
velocity it has been shown \cite{jay}, in the context of an $1-D$ 
Burgers equation, how $z=1$ could arise. \\

In this work we focus on the simplest two point non-equal-time velocity
correlation function in the wave-number ($k$) space.
We show that in $3-D$ fluid turbulence, in the large $L$ limit, 
the real part of the non-equal time velocity 
correlation function $C_{ij}({\bf k},t)\equiv
{\em R}\langle v_i({\bf k},t)v_j(-{\bf k},0)\rangle$, for the 
inertial scales ($l=k^{-1}$), has oscillatory behaviour within a
decaying envelope. From the incompressibility 
(${\bf \nabla . v}=0$) and isotropy assumptions it follows
$C_{ij}({\bf k},t)=c(k,t)P_{ij}({\bf k})$.
Here $P_{ij}({\bf k})=\delta_{ij} -k_i k_j/k^2$ is the tranverse 
projector. \\

An attempt to calculate the $c(k,t)$ has been carried out 
by L'vov et.al. \cite{lovov} in the context of a turbulent flow with a mean 
velocity field $V_0$. But they treated Navier-Stokes equation at a
linear level. They had compensated for the non-linear term to some 
extent by using a wave number dependent renormalised viscosity 
instead of the bare viscosity. They predict an oscillatory behaviour for
$c(k,t)$, but in the absence of the mean velocity 
(i.e., $V_0=0$) the oscillation vanish (i.e., it is a purely
kinematic effect). But our data from a numerical simulation of 
the $3-D$ NS equation (with large but finite $L$ and zero mean velocity) clearly 
reveals presence of oscillations. The data (see Fig.4,5) looks like 
the displacement of an under-damped harmonic oscillator. Also simulation of 
the REWA (reduced wave vector set approximation) model by Eggers 
\cite{rewa} shows a non exponential decay at short times and a 
negative minima. \\

Incompressible NS equation, forced randomly with a scale 
dependent variance, has been shown \cite{dedom,rfnse} to be a 
good model for fluid 
turbulence as far as multiscaling properties are concerned.
But there exist many unresolved theoretical problems with the 
analytic calculations with this model. Our calculation is based on 
a variant of this model, where instead of a singular forcing spectrum
(which goes as $k^{-3}$) we use a spectrum which peaks at a small but 
finite $L^{-1}$, goes to zero at $k=0$ and behaves as $k^{-3}$ for 
$k\gg L^{-1}$. The equation of motion
for the velocity field fourier component ${\bf v_i(k)}$ is
\begin{eqnarray}
\dot{v}_{i}({\bf k}) + \nu_0 k^{2}v_i ({\bf k}) =
&-i&\lambda M_{ijl} ({\bf k})\sum_{\bf q}v_{j}({\bf q}) v_{l}({\bf k-q})
\nonumber \\ &+& f_{i}({\bf k},t)\;.
\end{eqnarray}
The random force $f_i(k,t)$ is a gaussian, white noise with the variancne
\begin{equation}
\langle f_i({\bf{k}},t) f_j({\bf{k}}',t')\rangle =
\frac{(2\pi)^3 2D_0 k}{(k^2 + L^{-2})^2} P_{ij}({\bf k}) 
\delta({\bf k+k'})\delta(t-t')
\label{e.1} 
\end{equation}
Here  $M_{ijl}=[k_{j}P_{il}({\bf k}) + 
k_{l}P_{ij}({\bf k})]/2$ and $\lambda$ is an artificial coupling constant 
which will be set to $1$ later. Henceforth we will denote the variance of the 
force $2D_0k/(k^2 + L^{-2})^2$ by $D(k)$.\\

In the theories with singular forcing spectrum (or equivalently infinite 
integral scale), infrared divergences appear if one tries to calculate
effective viscosity perturbatively. Also the scheme cannot handle 
the so called sweeping effect i.e., the interaction
of the bigger eddies (of size $q^{-1}$) with the eddy of size $k^{-1}$
(when $(q<k)$). Physically the bigger eddies just advect 
the smaller eddies without distorting them much, so such divergences
are basically defect of such a perturbative scheme. But One should 
remember that this 
eddy picture is quite heuristic in nature because velocity fourier
modes $v(k)$ are global features of the velocity field where as the
eddies are spatially correlated patches in the velocity field and hence
local in nature. One systematic way to get rid of 
sweeping divergences in the equal time velocity structure functions, 
is to go to the lagrangian frame. Another way is to do an RG calculation
which excludes the effect of the $(q<k)$ modes on the $k$ mode, so 
both the infrared divergence and sweeping effect are eliminated. In
these calculations because of universality reasons one is mainly 
interested in the zero frequency limit of the effective viscosity 
i.e., $\delta \nu(k,\omega\rightarrow 0)$
and assumes that for all frequencies $\omega$, the effective propagator
$G(k,\omega)\sim (-i\omega + k^2 \delta \nu(k,\omega\rightarrow 0))^{-1}$ 
i.e., remains 
a Lorentzian in $\omega$. This approximation works well for long
time properties. But here since we are interested in $c(k,t)$ for all
$t$ (including the short time behaviour) we need the correct behaviour 
of $G(k,\omega)$ for all $\omega$s'. \\

Our procedure to calculate  $c(k,\omega)$ is a mixture of a self-consistent 
and a perturbative scheme. We show that if we assume a large but finite integral 
scale $L$, due to nonlinear interaction among modes, the effective 
viscosity is complex. It has the regular renormalized
real part and a $k,\omega$ dependent imaginary part as well. 
The oscillation in $c(k,t)$ arises because of this imaginary part. 
We use an one loop perturbation theory to determine
the complex viscosity. The calculation is similar to the standard 
RG procedure for evaluating zero frequency viscosity. But unlike in the
RG procedure, where only modes greater than the external wave-vector $k$ 
are integrated out,
we integrate over all $q$ modes, including the range $[0,k]$.
In the zero frequency limit ($\omega=0$) our calculation is 
self-consistent (at one loop level), but for finite $\omega$ it is
a perturbative calculation. Since our forcing spectrum is not singular,
there is no infrared divergence in our integrals. \\

Treating the nonlinear term perturbatively \cite{fns}
the effective response function $G(k,\omega)$ can be calculated as 
$G^{-1}= -i\omega + \nu _0 k^2 + \delta \nu (k,\omega)$. 
In the small $k$ and $\nu_0\rightarrow 0$ limit, $\nu _0 k^2$ is negligible 
compared to $\delta \nu (k,\omega)$. 
Hence $G^{-1}=-i[ \omega - k^2 {\em I(S)}] + k^2 {\em R(S)}$,
where we have denoted $\delta\nu (k,\omega)$  by $k^2 S$, and the real, 
imaginary parts by {\em R,I}. 
Using $c(k,\omega)=D(k)|G(k,\omega)|^2$ we get
\begin{equation}   
  \;\;\; c(k,\omega)=\frac{D(k)} {[\omega - k^2 {\em I(S)}]^2 +
k^4 {\em R(S)}^2}
\label{e.6}
\end{equation}

$c(k,t)$ is the inverse fourier transform of $c(k,\omega)$.
So our task is to calculate $S$. Following Ref.\cite{fns}

\begin{eqnarray}
-k^2 S P_{lj}({\bf k})&=& (i\lambda)^2 M_{lmn}({\bf k})
\int \frac{d^3 q d\omega'} {(2\pi)^4}  
M_{nij}({\bf k-q})\times \nonumber\\
& &P_{im}({\bf q})D(q)|G(q,\omega')|^2 G(|{\bf k-q}|,\omega-\omega')\nonumber\\
\label{e.3} 
\end{eqnarray}
Multiplying both sides by $P_{jl}({\bf k})$ and contracting over $l,j$
we get 
\begin{eqnarray}
S = \frac{M_{jmn}({\bf k})}{2k^2}&\int &
\frac{d^3 q d\omega'}{(2\pi)^4}M_{nij}({\bf k-q})
P_{im}({\bf q})D(q)\times\nonumber\\
& &|G(q,\omega')|^2 G(|{\bf k-q}|,\omega-\omega')\nonumber
\end{eqnarray}
Integrating the r.h.s. over $\omega'$ gives
\begin{eqnarray}
S=\int \frac{d^3 q} {(2\pi)^3} b({\bf k,k-q,q})
\frac{D(q)}{2\nu_1 q^z}  \frac{1}
{-i \omega + \nu_1 (q^z + |{\bf k-q}|^z) }
\label{e.4} 
\end{eqnarray}

$k^2 b({\bf k,k-q,q})$ is obtained by contracting $M_{jmn},M_{nij}$ 
and $P_{im}$. The expression for $b$ is $b({\bf k,k-q,q}) 
= \frac{|{\bf k-q}|}{k} (xy+z^3)$ \cite{lehus-orz}.
The trio $({\bf k,k-q,q})\/$ form a triangle and $x,y,z\/$
are the direction cosines of the angles opposite to
${\bf k,k-q}\/$, and ${\bf q}\/$ respectively. 
We have calculated this integral numerically.
We have used $G(k,\omega)^{-1}=-i\omega + \nu_1k^z$ 
anticipating a renormalisation of the viscosity in the 
$\omega\rightarrow 0$ limit. Here $\nu_1$ could 
be a function of $L,D_0$ which we determine later. But as far as 
the $k,\omega$ dependence of the integral is concerned, for fixed 
$L$ and $D_0$, we can treat $\nu_1$ as a constant. From Eq.5 note
that ${\em I(S)}\rightarrow 0$, as $\omega\rightarrow 0$. We check
numerically that at $\omega=0$, the integral in Eq.5 scales as 
$k^{-2z}$. Also by expanding the integral in small $k$ and 
retaining only leading powers in $k$ (as is done in a DRG 
procedure \cite{YOrg}) one can infer this. If self-consistency 
has to be achieved at $\omega=0$, then on the l.h.s. of Eq.5,
$R[S(k,0)]=\nu_1 k^{z-2}$. This fixes $z=2/3$.
Numerical evaluation of the integral with a finite $L$ corroborates
the approximate analytic prediction for small $k$ because $L^{-1}<<k$.  
Now we determine $\nu_1$ in a self consistent way \cite{jkb1}.\\

\begin{figure}
\epsfig{file=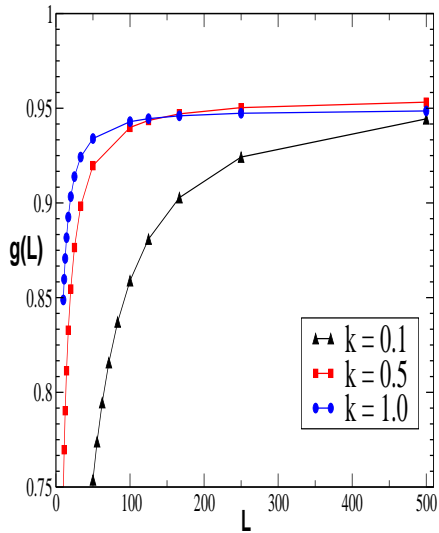, height=6.5cm, width=9cm, angle=-90}
\caption{$\nu_1(L)$ versus $L$. It shows convergence of $\nu_1(L)$ as
$L\rightarrow \infty$.
In our numerical evaluation of the integral in Eq.5 we choose $D_0=0.01$
and first evaluate the integral for $\omega=0$ in order to get Fig.1. From this 
and Eq.6 we determine $\nu_1\sim 0.06$. Then we choose a large enough
value for $L (=100)$ and evaluate the integral for nonzero values of
$\omega$ to get $S$.}
\end{figure}

At $\omega=0$, self consistency of Eq.5 requires

\begin{eqnarray}
\nu_1 k^{-4/3}=\frac{D_0}{(2\pi)^2 \nu_1^2} k^{-4/3}g(L)\;\;,\;\;
\mbox {so}\;\;
\nu_1^3 = \frac{D_0}{(2\pi)^2}g(L)\nonumber\\
\end{eqnarray}

Where $g(L)$ is the $L$ dependent integral on the r.h.s. of Eq.5.
In Fig.1 we show that $g(L)$ converges to a finite value as $L$ grows large. 
Using $g(L)\simeq .95$ from Fig.1, we get $\nu_1(L)\simeq 0.28 D_0^{1/3}$.
While evaluating the integral for inertial range $k$ modes we  
used $G^{-1} = - i\omega + \nu_1 k^{2/3}$ (neglecting the $\nu_0 k^2$ term) 
for all the $q$ and $k-q$ modes in the integrand, though the integral 
runs over both inertial and dissipation modes. But this is approximately 
correct because modes very 
far from the $k$ mode do not contribute (locality in $k$ space).\\

\begin{figure}
\epsfxsize=18pc
\epsfig{file=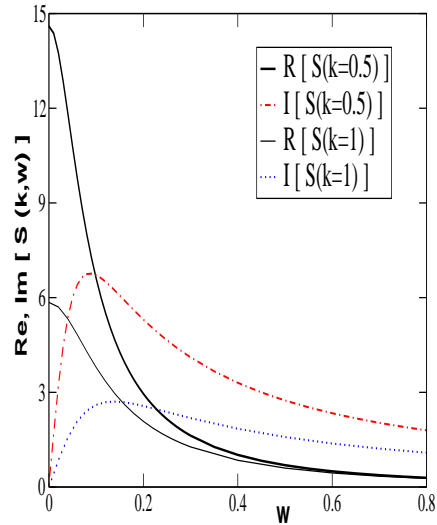, height=6.5cm, width=9cm, angle=-90}
\caption{$S$ versus $\omega$ for two different $k$ values. Real and imaginary
parts of $S$ are indicated by $R[S]$ and $I[S]$.}
\end{figure}

Returning to Eq.5, the dominant pole of $c(k,\omega)$ will decide the
oscillation and the decay of $c(k,t)$. If the pole lies at $\omega=
\omega_1 + i\omega_2$ then $\omega_1=k^2 {\em I(S}[\omega_1])$ and
$\omega_2=-k^2 {\em R(S}[\omega_2])$. From the shapes of 
${\em R,I(S)}$ in Fig.2 and the fact that ${\em R,I(S)}$ are
even,odd functions of $\omega$, we can infer that 
there will be two solutions to these transcendental equations. They 
are of the form $\pm \omega_1 + i\omega_2$, when $\omega_2$
is negative.  This ensures the causality of the effective response 
function $G$.  The approximate data collapse for the inertial range 
$k$ modes ($\eta_D^{-1}>k>L^{-1}$) in our Fig.3 implies that 
the oscillation period scales with $k^{-2/3}$.\\

\begin{figure}
\epsfig{file=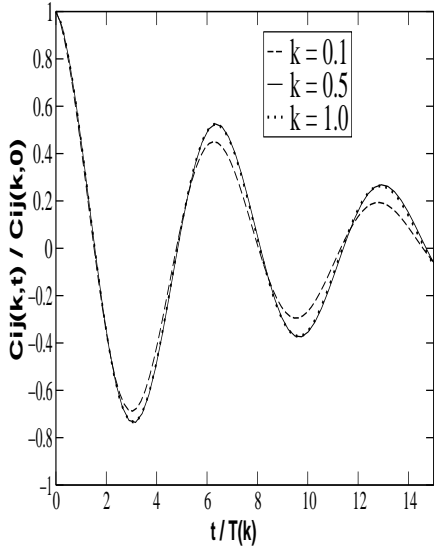, height=6.5cm, width=9cm, angle=-90}
\caption{$c(k,t)$ versus the scaled time $t/T(k)$ (with $T(k)=k^{-2/3}$),
for two different $k$ values.}
\end{figure}

We performed a DNS of the $3-D$ NS equation and calculated $c(k,t)$
versus $t$ for various ${\bf v(k)}$ modes.   
The data (see Fig.4,5) clearly shows oscillation in time. In our
simulation forcing was present only at large length
scales (i.e., in the $k$-space the ${\bf v(k)}$ modes in the smallest two 
shells were forced). Our pseudo-spectral scheme for the DNS is same as 
in \cite{menvin}. We used a $32^3$ grid with periodic boundary condition. 
We obtained a short inertial range (shown in Fig.6) 
with $Re_\lambda\sim 22$. We have obtained very long
time series ($\geq 68 T(L) $) for averaging $c(k,t)$. 
A preliminary test run \cite{dhrubo} 
on a $64^3$ grid also confirms existence of such oscillations. \\

\begin{figure}
\epsfig{file=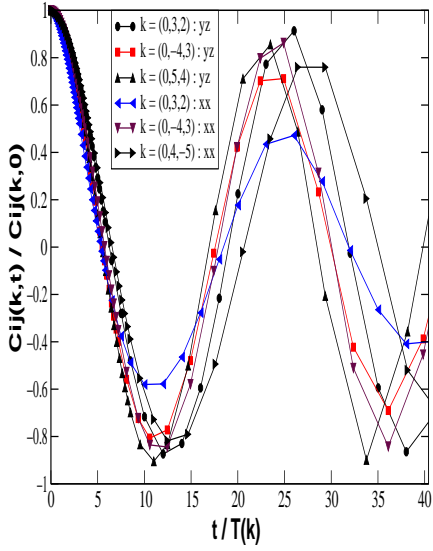, height=6.5cm, width=9cm, angle=-90}
\caption{$C_{ij}({\bf k},t)/C_{ij}({\bf k},0)$ plotted against
scaled time $t/T(k)$ (with eddy turnover time 
$T(k)=A\epsilon^{-1/3}k^{-2/3}$). Plots for different,
independent ${\bf k}$ vectors (in the inertial range) and different
$i,j$ are indicated by ${\bf k}=(k_x,k_y,k_z) : i,j$. Approximate
data collapse can be seen.  Data point shown at the longest time
separation  and zero time separation are averaged over $\sim 68$
and $\sim 84$ Large-eddy turnover times ($T(L)$) respectively. }
\end{figure}

\begin{figure}
\epsfig{file=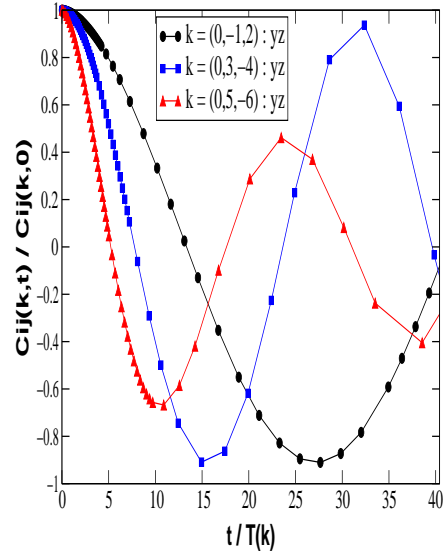, height=6.5cm, width=9cm, angle=-90}
\caption{$C_{ij}({\bf k},t)/C_{ij}({\bf k},0)$ versus $t/T(k)$ plotted for
three different ${\bf k}$ vectors which lie close to the forced range
(circle), in the inertial range (square) and close to the dissipation
range (triangle). They show wide separation in periodicity and decay.}
\end{figure}

\begin{figure}
\epsfig{file=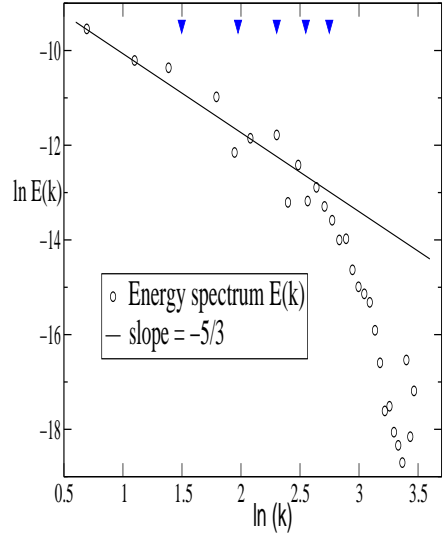, height=6.5cm, width=9cm, angle=-90}
\caption{The energy spectrum $ln E(k)$ versus $ln k$. The position of the
different chosen $k$ values, for which $c(k,t)$ has been plotted
in Fig.4 and 5, are indicated at the top of the figure.}
\end{figure}

Our simulation data in Fig.4 indicates that the oscillation 
time period approximately scales as $k^{-2/3}$ in the inertial range. 
But Fig.5 shows that modes closed to the forced 
modes and the dissipative modes differ widely from the inertial ones in
periodicity and decay. We have averaged the data for $\geq 68 T(L)$ 
(when $T(L)$ is the large eddy turnover time). This is
sufficienty long averaging time for an equal time, shell averaged, 
correlation functions to converge; but here for $c(k,t)$ since (a) single
${\bf k}$ mode is involved, (b) long time history (few $T(L)$'s) of 
the mode is important, it requires a longer averaging time.
In a simulation with small but finite $\nu_0$ the $\nu_1 k^{2/3}$ term 
will loose its dominance over $\nu_0 k^2$ as $k$ increases towards 
the dissipation range. That explains the strong damping seen in
the ${\bf k}=(0,5,-6)$ mode in Fig.5.\\

In the simulation we calculate $T(L)$ using the formula 
$T(L)=L_{box}/v_{rms}$, where $L_{box}$ is the simulation box size.
From dimensional analysis arguments \cite{frischbook} the scale dependent 
eddy-turnover time $T(k)=A\epsilon^{-1/3}k^{-2/3}$, when $A$ is a
constant of ${\cal O}(1)$.
Equating $T(L)=A\epsilon^{-1/3}(2\pi/L)^{-2/3}$ we get the prefactor 
$A\epsilon^{-1/3}$ and hence can calculate $T(k)$. Fig.4 shows 
that the time period of oscillation $\lambda_k\sim 25 T(k)$. This ratio may 
be non-universal. To get a clue let us look at our randomly forced model, where 
$I(S)$ is an explicit function of $D_0$, $\nu_1$ and $\nu_1\propto D_0^{1/3}$ 
(see Eq.5 and 6). Hence $\lambda_k$ depends on $D_0$ 
in a complicated way. We explore the dependence of $T(k)$ on $D_0$ below.
In this model $T(k)$ cannot be determined in a simple way because here
all the $k$ shells are being forced and hence the energy flux is not
a constant but increses logarithmically with $k$ \cite{rfnse}. 
The scale dependent energy flux $\Pi(k)=\int_0^k D(q)d^3q/(2\pi)^3=
\frac{D_0}{2\pi^2} [\ln(a^2 + k^2)- k^2(a^2 + k^2)^{-1}- 2\ln(a)]$,
when $a=L^{-1}$. Again using dimensional analysis we get
$T(k)=A.\Pi(k)^{-1/3}k^{-2/3}$ (hence $T(k)\propto D_0^{-1/3}$). 
Evaluating $\Pi(k)^{-1/3}$ for $D_0=0.01$ 
and $L=100$ (which we used for our theoretical graphs in Fig.2,3) 
gives $\Pi(0.1)^{-1/3}=6.17$, $\Pi(1.0)^{-1/3}=8.1$.
Allthough this $k$ dependence is weak, but 
the perfect data collapse with respect to $tk^{2/3}$ in  
Fig.3 implies that the above dimensional analysis estimate is not accurate.
In Fig.3 the oscillation period $\lambda_k \sim 6k^{-2/3}$ and hence
$\lambda_k\sim T(k)$ (neglecting the constant $A$ of ${\cal O}(1)$).  \\

The imaginary part of the viscosity, generated by an one loop perturbation 
theory, cannot be interpreted as a background 
velocity which is slowly varying in time. This mis-interpretation may be
provoked by the fact that if we had a mean background flow $V_0$
in the problem, then the nonlinear term would generate an extra term
$i({\bf V_0.k)v(k},t)$ in the NS equation. With this extra linear term the bare 
propagator will be $G(k,\omega)\sim (-i\omega  -i{\bf V_0.k} + 
\nu_0 k^2)^{-1}$ \cite{lovov}. 
But the $k,\omega$ dependent complex viscosity, which we find in our theory,
when transformed to $({\bf x},t)$ space, gives an additional
memory term $\int d{\bf x'}dt'\kappa({\bf |x-x'|},t-t'){\bf v(x'},t')$ 
in the equation of motion (e.o.m.). 
But it is true that the $\kappa(k,t)$ field oscillates
at a slower time scale than that of the ${\bf v(k},t)$ mode itself 
(as ${\em I}\delta\nu (k,\omega)$ has a peak at a lower $\omega$
than that of $c(k,\omega)$).
Also a significant contribution to the integral for $\delta\nu (k,\omega)$
comes from the $q<k$ modes. So it does resemble sweeping by larger (and hence 
slower) eddies to some extent. But rigorously convection by a
background velocity ${\bf V_0(x},t)$ should look like 
${\bf V_0(x},t).{\bf\nabla v(x},t)$ (which is local in $({\bf x},t)$).
We note that in the turbulence context, complex effective viscosity 
has been proposed before by 
J.K. Bhattacharjee in \cite{jkb2}. The author had assumed
dynamic scaling hypothesis (as in dynamic critical phenomena)
to be valid for fluid turbulence and had predicted a form for the effective
viscosity in the high frequency limit. Then an interpolation scheme had 
been used to connect the two limits (small $\omega$ and large $\omega$)
of the effective viscosity.\\

In conclusion, we have given numerical evidence and an approximate theory
for the novel oscillatory decay of the two-point, temporal correlation 
function in $3-D$ fluid turbulence. This behaviour is similar to 
viscoelastic effect seen in complex fluids. \\

We would like to thank J.K. Bhattacharjee (of IACS Calcutta, India)
for some critical comments and R. Pandit, S. Ramaswamy (of IISc 
Bangalore, India) and M. Plischke (of SFU, Canada) 
for discussions. Also thanks to SERC,IISc Bangalore, 
(India) for computational facilities and NSERC (Canada) 
for supporting work in Canada.

\end{document}